\documentclass[twocolumn]{aastex701}
\usepackage{hyperref}
\usepackage{caption,subcaption}
\usepackage{tabularx}
\usepackage{amsmath}

\begin{document}

\title{On-sky demonstration of second-stage wavefront control with a photonic lantern}

\author[0000-0002-5669-035X]{Aditya R. Sengupta}
\affiliation{Department of Astronomy and Astrophysics, University of California, Santa Cruz, CA 95064, USA}
\email[show]{adityars@ucsc.edu}  
\author[0000-0001-8786-1329]{Jordan Diaz} 
\affiliation{Department of Astronomy and Astrophysics, University of California, Santa Cruz, CA 95064, USA}
\email{}

\author[0009-0003-2280-9820]{Matthew DeMartino} 
\affiliation{Department of Astronomy and Astrophysics, University of California, Santa Cruz, CA 95064, USA}
\email{}

\author[0000-0003-0054-2953]{Rebecca Jensen-Clem} 
\affiliation{Department of Astronomy and Astrophysics, University of California, Santa Cruz, CA 95064, USA}
\email{}

\author{Sylvain Cetre} 
\affiliation{Wakea Consulting}
\email{}

\author[0000-0002-3739-0423]{Elinor Gates} 
\affiliation{University of California Observatories/Lick Observatory, PO Box 85, Mount Hamilton CA, 95140, USA}
\email{}

\author[0000-0001-9742-3138]{Kevin Bundy} 
\affiliation{Department of Astronomy and Astrophysics, University of California, Santa Cruz, CA 95064, USA}
\email{}

\author{Daren Dillon} 
\affiliation{University of California Observatories, UC Santa Cruz, 1156 High Street, Santa Cruz, CA 95064, USA}
\email{}

\author[0000-0002-1954-4564]{Philip Hinz} 
\affiliation{University of California Observatories, UC Santa Cruz, 1156 High Street, Santa Cruz, CA 95064, USA}
\email{}

\author[0000-0002-5082-6332]{Ma\"{i}ssa Salama} 
\affiliation{Department of Astronomy and Astrophysics, University of California, Santa Cruz, CA 95064, USA}
\email{}

\author[0000-0002-9372-5056]{Nour Skaf} 
\affiliation{Institute for Astronomy, University of Hawai’i at Manoa, Hilo, HI 96720-2700, US}
\email{nskaf@hawaii.edu}

\author[0000-0002-1097-9908]{Olivier Guyon}
 \affiliation{Subaru Telescope, National Astronomical Observatory of Japan,
650 North A`oh$\bar{o}$k$\bar{u}$ Place, Hilo, HI  96720, USA}
\affiliation{Astrobiology Center, 2-21-1 Osawa, Mitaka, Tokyo 181-8588, Japan}
\affiliation{Wyant College of Optical Sciences, The University of Arizona, 1630 E University Boulevard, Tucson, AZ, USA}
\email[]{guyon@naoj.org}

\author{Tara Crowe} 
\affiliation{CREOL, the College of Optics and Photonics, University of Central Florida, 4000 Central Florida Blvd, Orlando, FL 32816}
\email{}

\author[0009-0000-3311-3165]{Caleb Dobias} 
\affiliation{CREOL, the College of Optics and Photonics, University of Central Florida, 4000 Central Florida Blvd, Orlando, FL 32816}
\email{}

\author[0000-0002-1332-5061]{Stephen S. Eikenberry} 
\affiliation{CREOL, the College of Optics and Photonics, University of Central Florida, 4000 Central Florida Blvd, Orlando, FL 32816}
\email{}

\author[0009-0007-8546-4363]{Rodrigo Amezcua-Correa} 
\affiliation{CREOL, the College of Optics and Photonics, University of Central Florida, 4000 Central Florida Blvd, Orlando, FL 32816}
\email{}

\author[0000-0002-6268-3674]{Stephanos Yerolatsitis} 
\affiliation{CREOL, the College of Optics and Photonics, University of Central Florida, 4000 Central Florida Blvd, Orlando, FL 32816}
\email{}

\begin{abstract}
    Ground-based direct imaging of exoplanets at high contrast requires precise correction of atmospheric turbulence using adaptive optics (AO). The planet-to-star contrast ratio at small angular separations from the host star is often limited by non-common-path aberrations (NCPAs) seen only in the science plane. The photonic lantern (PL) can be used to sense aberrations at the final science imaging plane. This enables a two-stage wavefront control architecture, in which the first-stage wavefront sensor senses atmospheric turbulence and the PL senses NCPAs and other aberrations not seen by the first stage. We demonstrate closed-loop control of residual wavefront errors using a non-dispersed PL after first-stage AO correction on the Shane 3m telescope at Lick Observatory. Our results show that non-dispersed PLs can be used for second-stage wavefront sensing, enabling performance improvements via minimally invasive retrofits to existing AO systems.
\end{abstract}


\section{Introduction}

Direct imaging of exoplanets at high contrast using large ground-based telescopes requires advanced adaptive optics (AO) systems to counteract the effects of atmospheric turbulence and instrumental aberrations. AO systems measure aberrations in real time with a wavefront sensor (WFS) and apply corrections by changing the shape of a deformable mirror (DM). Non-common path aberrations (NCPAs), aberrations arising only in the path of the science instrument, are an important limiting factor to AO performance because they cannot be seen by the WFS. NCPAs with low spatial orders are a major limiting factor to achieving high contrast at small planet-to-star separations; they cannot be fully removed by off-sky point spread function (PSF) sharpening routines because they vary on timescales of seconds to hours (e.g. \citealt{Milli16,Vigan19}). NCPAs can also be addressed using on-sky calibrations; for example, \cite{Skaf22} demonstrated on-sky image quality improvements using a lucky-imaging-based approach. However, that approach required a high-speed focal-plane camera. This may not be compatible with science detector requirements, and separating such a high-speed camera from the science camera would introduce new NCPAs. Sensing aberrations in the plane of the final science image -- an example of a technique called focal plane wavefront sensing -- provides an opportunity to correct NCPAs in real time and improve imaging quality.

Photonic lanterns (PLs) are focal-plane wavefront sensors that are particularly suited for exoplanet direct imaging, as they can be used for wavefront sensing at the science plane while directing light downstream to other instruments. PLs are optical waveguides in which many single-mode fibers taper to a  single multimode waveguide, such as a multimode fiber \citep{Birks2015}. Light that comes to a focus at the PL input excites each mode by a different amount depending on the incoming phase. This gets mapped into intensity variations across the output single-mode fibers, which can be used as a WFS signal \citep{Corrigan2018,Norris2020,Wright2022}. At a single wavelength, the number of spatial modes PLs can sense is limited by their number of output ports \citep{Lin2023sim}. The PLs that have been tested for astronomy so far have had between 6 and 19 ports, and hence are well-matched to the low-order wavefront sensing application. PLs have been demonstrated to work as focal-plane wavefront sensors in laboratory settings \citep{Norris2020,Sengupta2024,Wei24,Xin24}. Relative to other focal-plane WFSs, they are compact and easy to install, as they can be connectorized like other fibers. This makes them potentially suitable for retrofits to existing ground-based telescopes or as the primary WFS on space-based coronagraphic imaging missions.

Linear wavefront reconstruction with the PL shows relatively small dynamic ranges \citep{Lin22,Sengupta2024} compared to the full scale of atmospheric turbulence. While several nonlinear reconstruction methods have been proposed and some have been demonstrated in the lab (e.g. \citealt{Wei24}), they are difficult to implement and currently require thousands to tens of thousands of examples for model calibration \citep{Norris2020,Lin_Fitzgerald_2024}. Therefore, in order to unify wavefront sensing and science imaging on one path while accommodating dynamic range restrictions, PLs need to work together with traditional pupil-plane WFSs in a two-stage configuration, in which aberrations with high spatial and temporal frequencies are corrected using the first WFS and those with low spatial and temporal frequencies are corrected using the PL. Second-stage wavefront control with a 19-port photonic lantern was recently demonstrated at Subaru/SCExAO \citep{Lin2023, Lin2025}, using dispersed PL outputs over a wide spectral range (1-1.8 $\mu$m). The light from each output port was dispersed using a prism to produce a spectrum per port, which was used as the WFS signal; this multi-wavelength information enabled 52 control modes.

The goal of this work is to investigate the performance of an undispersed PL over a narrow spectral range (1550 $\pm$ 30 nm). Demonstrating closed-loop control in this configuration would enable more light to be directed to other science instruments downstream using a simple optical design, while maintaining high image quality by correcting the leading-order non-common-path aberration modes.

In this work, we demonstrate closed-loop control, i.e. simultaneous improvement in PSF quality and reduction in measured wavefront error at the science imaging plane, using the photonic lantern in a dual-wavefront sensor single-conjugate configuration to correct static/slow non-common-path errors. We installed a 19-port PL on the Shane 3m telescope at Lick Observatory on Mount Hamilton, California. We calibrated the PL's response to aberrations induced by the Shane DM, and used this model to update the DM position in real time to improve the quality of the science image. We observe an improvement in peak PSF count of 15\% in closed loop relative to open loop.

The remainder of this paper is structured as follows. Section 2 describes the experimental setup, including optics for the PL input and output and the software pipeline. Section 3 describes the AO calibrations and observations. Section 4 shows results from on-sky wavefront control. Section 5 is a conclusion and a discussion of future work. 

\section{Experimental setup}

\begin{figure*}
    \centering
    \includegraphics[width=\linewidth]{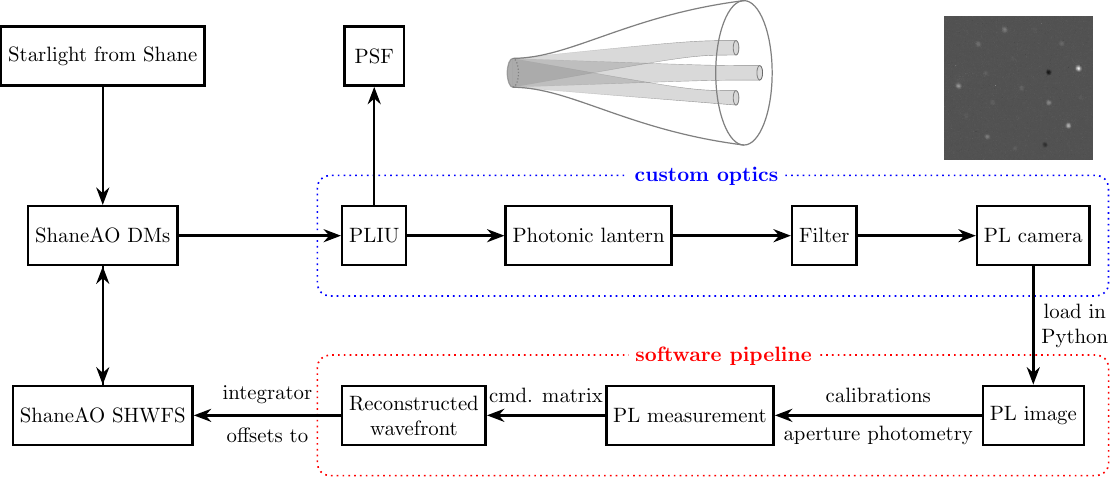}
    \caption{The experimental setup. The middle part of the diagram (in the blue box) leading from the ShaneAO DMs describes the optical setup installed for this work, and the lower part (in the red box) describes the software pipeline that was used. A representative PL image from this setup is shown in the top right.}
    \label{fig:control_configuration}
\end{figure*}

Figure~\ref{fig:control_configuration} shows the overall experimental setup used in this work, described in detail in the following subsections.

\subsection{The photonic lantern}
\label{sec:pl}

The photonic lantern used in this experiment was manufactured at CREOL, the College of Optics and Photonics at the University of Central Florida. It has 19 ports and has a design wavelength of 1550 nm; we filter the output to 1550$\pm$30 nm. This ensures that no light past the single-mode cutoff wavelength propagates through the lantern, meaning we observe a single mode per port so as to get unique wavefront-dependent patterns. Based on microscope images, we measure an input core diameter of 28 $\mu$m. This corresponds to $V = 7.37$, which would allow the PL to support around 15 modes. This is sufficient for the experiment described in this work.

\subsection{Optical setup and integration with ShaneAO}
\label{sec:optics}
We mount the PL input using the Parallel Lantern Injection Unit (PLIU), a custom assembly that picks off AO-corrected light on the Shane telescope and directs it to PLs and other devices. This assembly was described in \cite{DeMartino2022} and has served a prototype to the Astrophotonics Advancement at Lick Observatory (APALO) platform currently under development. As a prototype, the PLIU suffers from aberrations and flexure but offers sufficient performance for the PL tests we describe here. 

\begin{figure*}
    \centering
    \includegraphics[width=\linewidth]
    {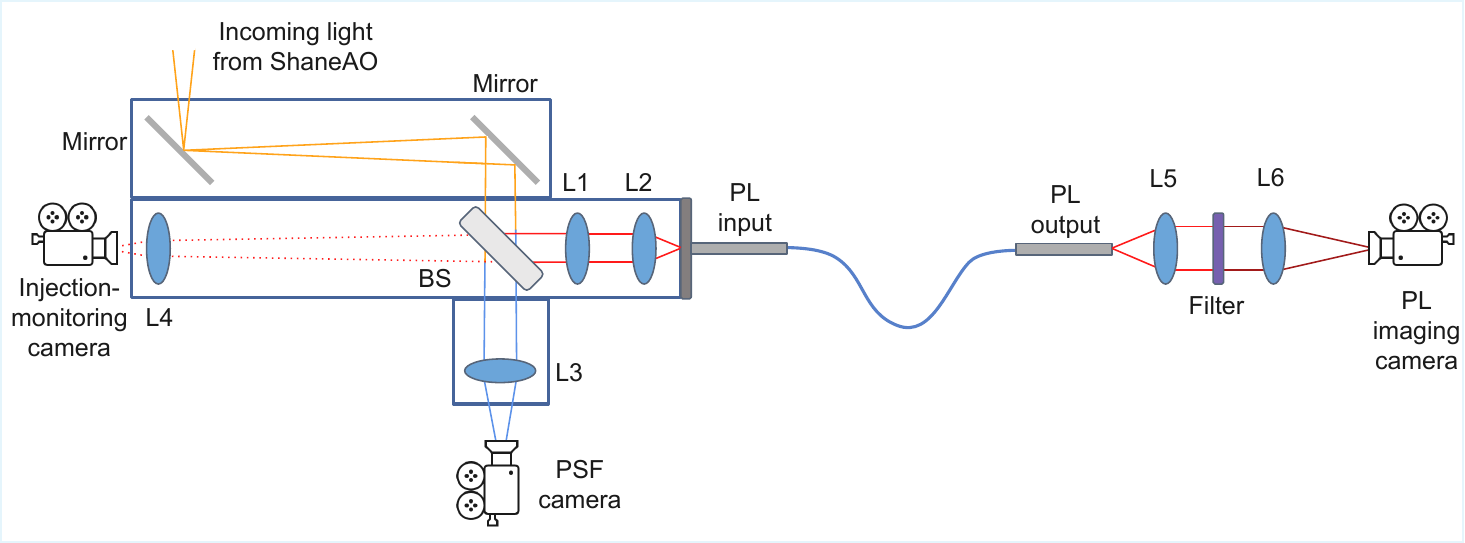}
    \caption{The optical layout of the PLIU and output imaging configuration used in this experiment. The dotted line from the beam splitter to the injection-monitoring camera is a back-reflection from the PL input.}
    \label{fig:optical_layout}
\end{figure*}

\begin{figure*}
    \begin{subfigure}{0.26\linewidth}
        \centering
        \includegraphics[width=\linewidth]
        {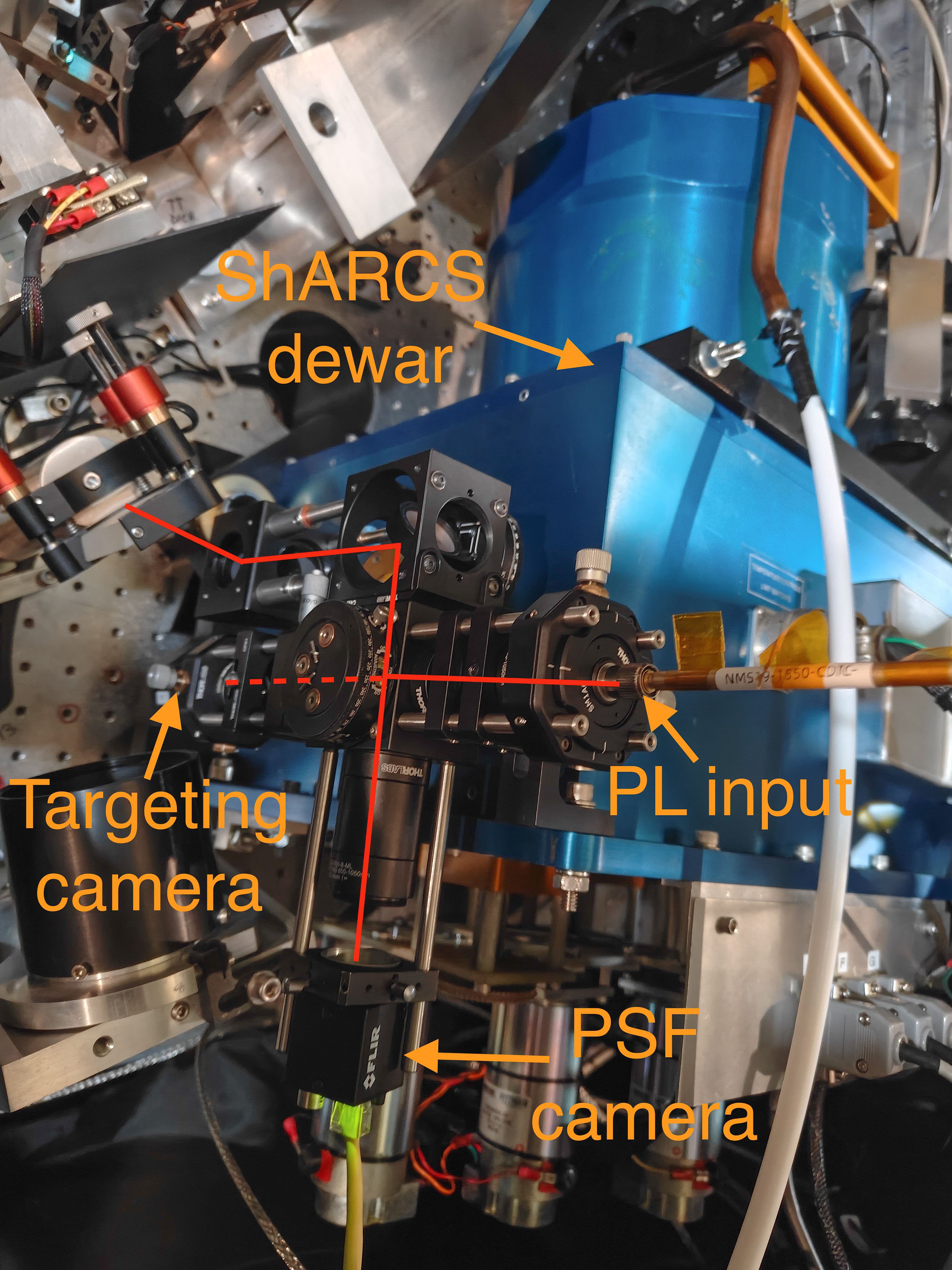}
        \caption{The PLIU/the lantern input.} \label{fig:pl_input}
    \end{subfigure}\hspace*{\fill}
    \begin{subfigure}{0.46\linewidth}
        \centering
        \includegraphics[width=\linewidth]
        {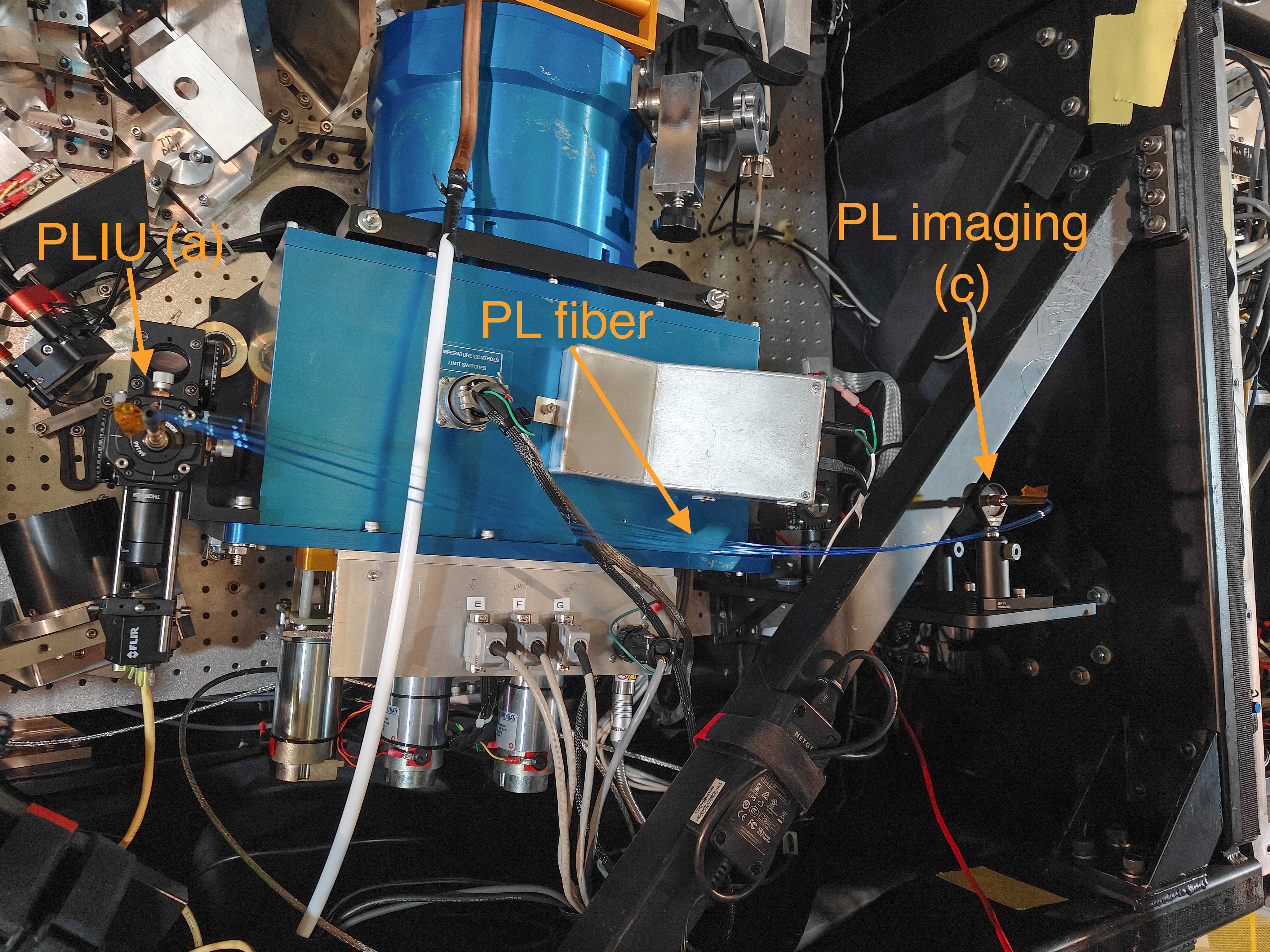}
        \caption{The PL fiber relative to ShARCS.} \label{fig:pl_fiber}
    \end{subfigure}\hspace*{\fill}
    \begin{subfigure}{0.26\linewidth}
        \centering
        \includegraphics[width=\linewidth]{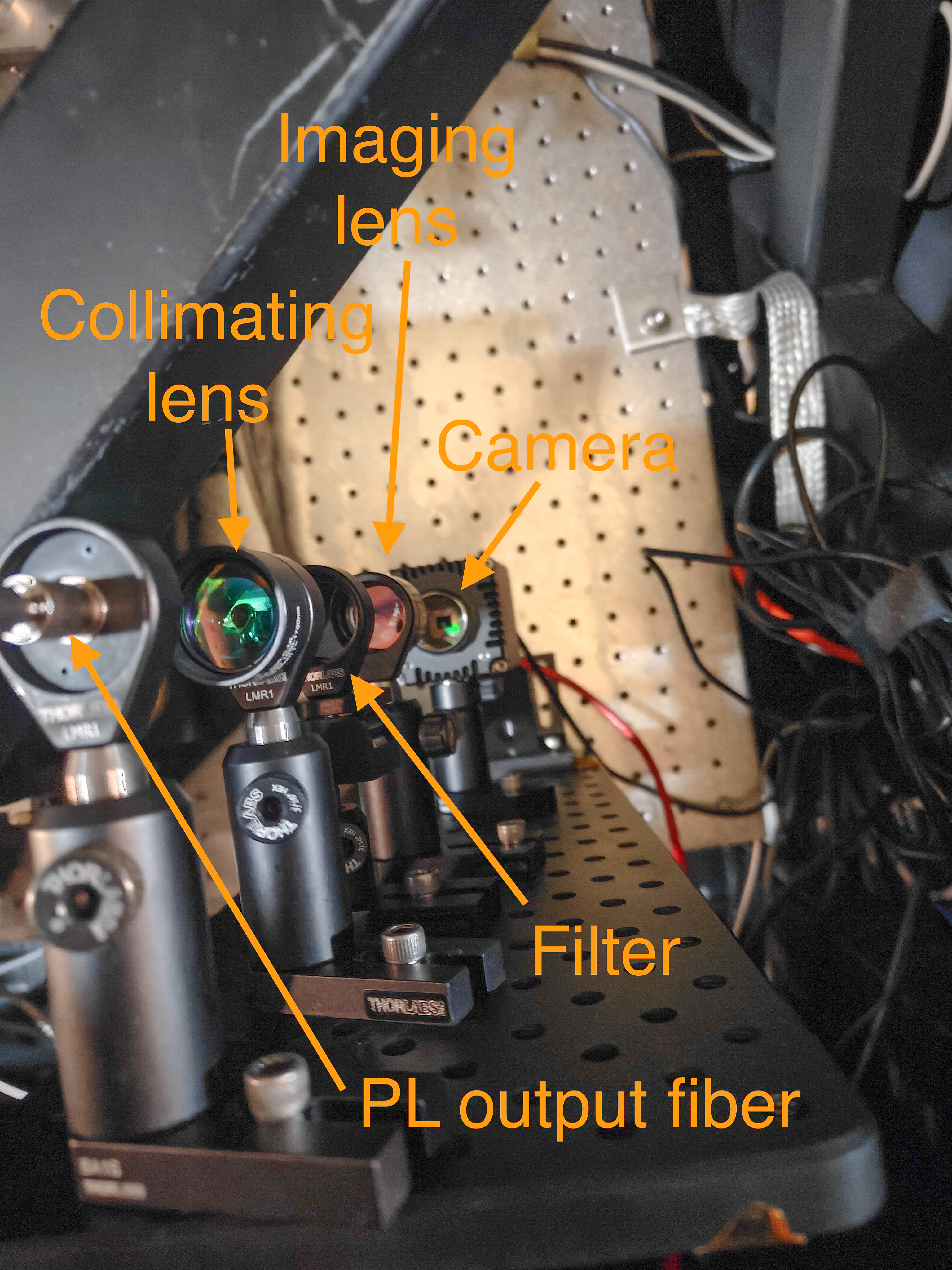}
        \caption{The PL output bench.}
        \label{fig:pl_output}
    \end{subfigure}
    \caption{The PLIU installed within ShaneAO. Light coming into the PLIU is converging and comes to a focus on a mirror, diverges as it passes through a beam splitter, and is collimated and focused along three separate paths.}
    \label{fig:pliu}
\end{figure*}

Light from ShaneAO goes through a 900+nm dichroic and is then injected into the PLIU. The PLIU mounts to an outer flange on ShARCS, the AO-fed near-IR camera \citep{McGurk2014}, replacing a filter wheel that normally sits in front of the ShARCS dewar window.  The converging beam is thus intercepted and directed instead into the PLIU's optical system, which includes fold mirrors and beam-splitters that guide the AO-corrected beam into the PL while providing for an injection-monitoring camera (a Blackfly U3-23S6M-C) to capture back-reflected light of{f the} PL's input facet.  This helps us to align and focus the system. For this experiment, the output of the PL---a loosely held bundle of 19 SMFs---was suspended over and across the ShARCS dewar so that the PL's output ferrule, which holds the 19 SMFs in a close-packed hexagonal array, could be mounted onto the PLIU's analysis arm.  This let us mount the analysis arm on available bench space which afforded more room for our camera system, given the tight space constraints around the PLIU's main assembly.  Figure~\ref{fig:optical_layout} shows the optical layout and Figure~\ref{fig:pliu} shows the installed configuration of the PLIU and the output imaging arm within the Shane optical bench. The PSF imaging arm was newly installed (since \citealt{DeMartino2024}) for this experiment.

\begin{figure}
    \centering
    \begin{subfigure}{0.5\linewidth}
        \centering
        \includegraphics[width=\linewidth]
        {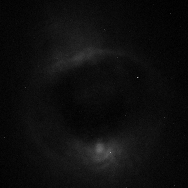}
        \caption{Inj.-monitoring camera} \label{fig:targeting}
    \end{subfigure}\hspace*{\fill}
    \begin{subfigure}{0.5\linewidth}
        \centering
        \includegraphics[width=\linewidth]{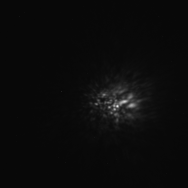}
        \caption{PSF camera}
        \label{fig:psfcam}
    \end{subfigure}
    \caption{Outputs from the two aligning cameras at the best alignment. We expect to see a faint ring with no light in its center on the injection-monitoring camera when well aligned.}
    \label{fig:cameras_align}
\end{figure}

When injected into the PLIU, light reflects off of a mirror and then encounters a 90:10 beam splitter that sends the majority of the beam to the PL but reserves a portion for the PSF-viewing camera (also a Blackfly U3-23S6M-C). Figure~\ref{fig:cameras_align} shows representative outputs from the injection-monitoring camera and PSF camera. Figure~\ref{fig:targeting} shows the back-reflection from the input fiber. When well aligned, we observe a faint ring-like structure from the wings of the PSF spilling beyond the PL's fiber face and reflecting off the PL's cladding and mounting ferrule; most of the PSF itself is not visible as it gets coupled into the PL, meaning that the majority of light transmits through the PL to the output stage when good alignment is achieved. A misalignment causes us to see brighter portions of the PSF as well as the ring from the lantern mount (Figure~\ref{fig:targeting}) on the injection-monitoring camera, and much less light through the PL. 

As a result of the 900+nm dichroic and both cameras' sensitivity out to 1000 nm, the PSF and injection-monitoring cameras see light in the 900-1000 nm range, { which} differs from the PL wavelength of 1550 nm. { Additionally, statically removing instrumental errors that are non-common-path between the PSF and PL in order to maximize PL throughput may result in additional error only in the PSF imaging path. These factors lead} to visible aberrations on the PSF when optimizing for PL throughput (see \S\ref{sec:coupling}). 

The final lens in the PLIU { (L2 in Figure~\ref{fig:optical_layout})} before the PL has a focal length of 25 mm. { The beam from ShaneAO is intercepted by the PLIU as a slowly converging beam, because it picks off light that would otherwise come to a focus on the ShARCS detector. We use the beam diameter within ShaneAO before injection into the PLIU, which is $D = 2.73$mm, as our reference for the size of the PSF being injected into the photonic lantern. We therefore compute a PSF core size of $ 25 \text{mm} \times 1.55 \mu \text{m} / 2.73 \text{mm} = 14.2 \mu$m.} This allows PSF features out to around $2 \lambda/D$ to fall within the MMF core. The numerical aperture of the PL is 0.11, which corresponds to an f/4.54 beam. Since the beam from ShaneAO has an f-number of 9.16, it is sufficiently slow that no light is lost on the sides of the multimode end. Nevertheless, the coupling is significantly suboptimal. Throughput is further limited by the PLIU's aberrations (discussed further in \S\ref{sec:conclusion_future}). We expect that higher throughput and/or { more significant responses to wavefront errors} will be achievable with an improved optical design.

In the analysis arm, the PL's output passes through a narrowband filter centered at 1550 nm with a bandwidth of 30 nm. Previous attempts using a broad H-band filter yielded no significant { variation in the distribution of light across the PL ports} as a function of wavefront aberrations. Following this, to test narrower filters, we made use of the muirSEAL testbed \citep{Sengupta2025}, which is an optical bench designed to test wavefront reconstruction with PLs at 1550 nm. We injected light from a white light source into this PL and filtered the output using the 30nm filter, and successfully demonstrated wavefront reconstruction, validating the use of this filter for the on-sky setup.

After passing through the filter, the PL output focuses onto a Goldeye G-030 VSWIR TEC1, the same detector used on the muirSEAL testbed. This is an InGaAs detector with a wavelength range of 900-1700 nm. The images have 520$\times$656 pixels, and the temporal dark noise at our gain level is 210 $\text{e}^-$. 

\subsection{Software setup}
\label{sec:software}
The Goldeye detector is read out and the images are loaded into Python in real time using the Durham Adaptive Optics (DAO) real-time software \citep{dao,Skaf24} onto a 2021 M1 MacBook Pro (hereafter the `control computer'). The PSF output is read out in the SpinView GUI on a Microsoft Surface laptop. Frames are automatically saved with timestamps accurate to the second and synchronized with the Goldeye frames in postprocessing; this is sufficiently accurate considering the mean DM latency of $\sim$1s. Direct connections over Ethernet are possible due to a custom-installed dedicated network switch with two Ethernet ports running from the Telescope Utility Bin \citep{McGurk2014} to the readout room.

We initially identify PL spots in the Goldeye image manually, and refine the positions by taking a center of mass around each spot. We make masks for aperture photometry by choosing an aperture size for all the ports that maximizes SNR over a time-series of 15 images captured on the ShaneAO calibration source. To compute SNR, we first add together the counts through each of the 19 apertures, giving us one measurement for each of the 15 frames. We then divide the mean of these measurements by the standard deviation.

To generate reduced measurements for wavefront sensing, we subtract a dark frame and sum the counts through each aperture. We set any negative outliers (below a few hundred negative counts), which correspond to no light coming through that port and the dark frame having significantly more intensity than that particular new frame, to zero. Removing these negative outliers was necessary when it was not possible to update the dark frame, or when a new dark frame showed significantly higher counts than previous measurements. This is possibly a result of the Goldeye detector overheating, which we observed after 1-2 hours of continuous operation. This caused higher dark counts; the on-board temperature sensor read 35${}^\circ$C at the end of daytime calibrations, up from the ambient temperature at the time of 20${}^\circ$C. We remedy this by updating the dark frame regularly, and on one occasion by turning off the camera for a 15-minute interval to allow it to cool. We sweep through darks for each candidate exposure time whenever an update is required, in case improved or degraded alignment requires us to change the exposure time. 

After dark subtraction, we divide all resulting entries by the sum of all aperture counts in the reduced frame. This normalization is necessary to avoid any wavefront reconstruction dependence on either source brightness or exposure time. This enables model calibration on the brighter internal source for testing on-sky with a dimmer target on which we see larger aberrations.

\section{Adaptive optics calibration}

\subsection{Initial coupling and deformable mirror interfacing}
\label{sec:coupling}
We use the PL to control ShaneAO's two deformable mirrors: the high-stroke low-order `woofer', and the low-stroke high-order `tweeter' \citep{Gavel2014}. We interface with the DMs by sending reference offsets for the Shack-Hartmann wavefront sensor. Since we only move in low orders, the woofer primarily moves in response, but due to the two-DM offloading control scheme used by ShaneAO, the tweeter may also move. We send commands via an image sharpening interface designed for observers using ShARCS. This is ordinarily a GUI intended to correct static aberrations by eye on the ShARCS science camera.  The interface allows us to correct 12 Zernike modes: focus, two astigmatism modes, four 3rd order modes, and five 4th order modes. A bash version of this script was made available to us over SSH. We maintain a persistent SSH connection from the control computer to the ShaneAO real-time computer using the \textit{paramiko} Python library \citep{paramiko}. 

\begin{figure}
    \centering
    \includegraphics[width=\linewidth]{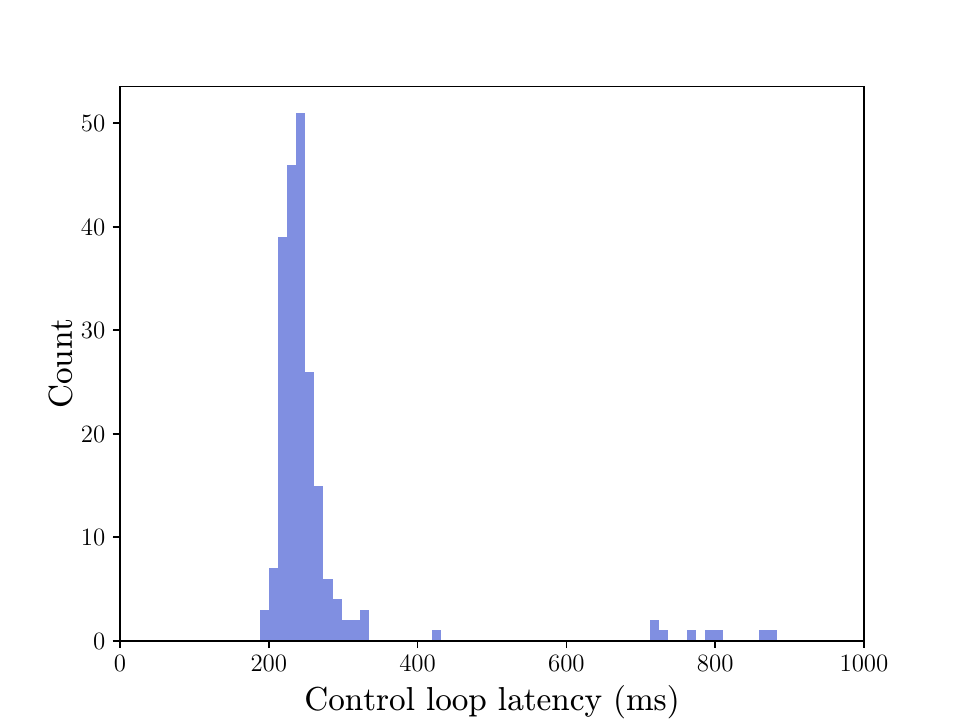}
    \caption{Latency of the control loop as measured on the control computer. Three measurements were taken above 1000 ms; for clarity, these are not shown.}
    \label{fig:loop_latency}
\end{figure}

Communication over Wi-Fi together with the inherent latency of the image sharpening script yielded an overall DM latency of $\sim$1s. Since timestamps for DM updates could not be sent back from ShaneAO to the control computer, a more precise assessment of the DM latency could not be made. However, we are able to assess latency on the control computer by subtracting the WFS exposure time from the mean time between DM updates. { Leaving out 12 outlier measurements past 400 ms, w}e measure a latency of { 242 ms $\pm$ 23 ms. We also note that the high-latency outliers were non-consecutive, were likely caused by momentarily higher network load, and were unlikely to significantly alter control results.} Figure~\ref{fig:loop_latency} shows the distribution of latencies on the control computer, which includes computation of the control command and sending it to ShaneAO. { This latency is well matched to the timescale on which quasi-static non-common-path aberrations evolve, but would not be sufficient to correct residual atmospheric aberrations, which evolve on millisecond timescales (set by the ShaneAO loop rate of around 1 kHz). Since the typical exposure time required on sky ($\sim1-3$s) already limits our ability to correct atmospheric residuals, our experiment is focused on quasi-static aberrations and therefore our system latency is only required to be less than roughly half a second. The measured latency is therefore acceptable for our purposes.}


When acquiring an initial image { off sky}, we install the PLIU and sharpen the PLIU's PSF image by eye out to the 4th order modes. Next, we manually adjust the tip, tilt, and focus of injection into the PL to maximize the counts at the PL output to achieve a best DM flat position. Quasi-static mechanical drift from the AO system can cause a loss of tip/tilt precision over time; further, there is no direct access to tip/tilt on the woofer available over SSH. This means that over time the initial coupling during alignment becomes difficult to achieve again. 

Since PL throughput and wavefront reconstruction quality are highly sensitive to tip/tilt offsets, we optimize tip/tilt in open loop before closing the loop on the other orders. We optimize tip/tilt in open-loop via a spiral search through DM positions. This goes through a grid of tip/tilt positions centered on the current zero, moving by one step in either the north/south or the east/west direction each time. The spiral search pattern means there are no discontinuous jumps in throughput compared to a linear search; on sky, this allows us to discriminate between a true throughput peak due to tip/tilt optimizing and a temporary throughput improvement due to improved conditions coinciding with one particular position. At each point in the spiral, the script prints the timestamp and current coordinates and waits for a fixed number of seconds, set manually to allow two to three exposures on the PL camera per position. { Depending on the target, conditions, and alignment quality, the exposure time on the PL camera may range from 1-5s, leading us to choose wait times for the spiral search in the range 2-20s.} Simultaneously, we run a script on the control computer that repeatedly reads out the PL camera and prints out the sum of counts through all ports. We manually match the timestamp of the highest PL throughput to the DM position and set the new best flat to that position. This procedure {is} typically run 2-3 times, starting with coarse step sizes between 0.06''-0.1'' and later reducing this to a finer step size of 0.01''. { This is comparable to or smaller than the diffraction-limited PSF size on sky, which is roughly 0.1''.} Empirically, this optimization needs to be redone roughly once per hour when observing on sky or once per two to three hours of testing with the internal ShaneAO calibration source.

{ During alignment, we observe} some degradation in PSF quality when steering in tip/tilt, requiring some additional image sharpening at the new position. { The source of this coupling between modes is unclear.} 

{ In order to obtain direct evidence of improved image quality as a result of second-stage correction, we chose to measure PSF quality as our main figure of merit. This comes at the cost of non-common-path aberrations between the PSF camera and the PL, but these are relatively small relative to those that we are attempting to correct, which exist between ShaneAO's first-stage WFS and the PL. However, we are still required to optimize coupling into the PL to ensure a high-quality signal for wavefront sensing, and it has not proven to be possible to maximize PL throughput while also achieving the best PSF position and shape. In addition to optical defects, this issue is likely affected by the wavelength difference between the PL camera and the PSF and injection-monitoring cameras. For this experiment, w}e empirically choose a best flat position that is close to the optimum for both the PL and the PSF. Revisions to the optical design may be able to provide an improved joint optimization (see \S\ref{sec:conclusion_future}).

\subsection{Interaction and command matrices}

\begin{figure}
    \centering
    \includegraphics[width=\linewidth]{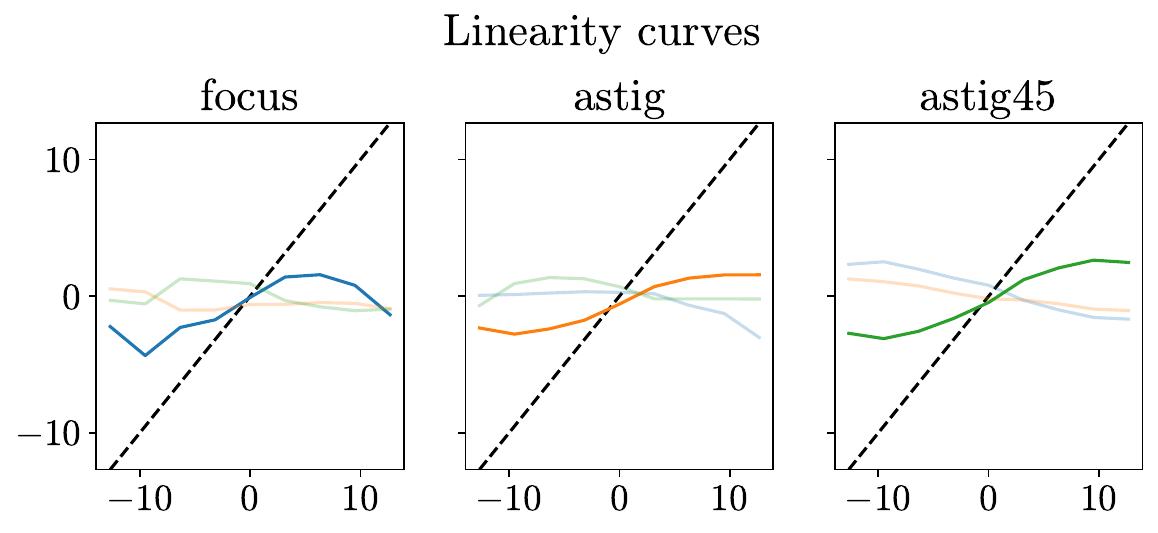}
    \caption{Linearity curves in focus/astig/astig45 on the ShaneAO calibration source. Both axes are in nanometers.}
    \label{fig:linearity}
\end{figure}

We measure an interaction matrix $M$ in the Zernike modal basis set using the image sharpening script. { This measurement is carried out off sky.} We follow the usual procedure for AO calibration, by measuring positive and negative pokes in each mode and collecting the resulting differential signals on the PL as the columns of the interaction matrix. We used a poke amplitude of 1.268 nm, corresponding to 0.002 in ShaneAO's image sharpening units (see \S\ref{sec:image_cal}). The poke amplitude was chosen empirically, by assessing linear ranges (shown in Figure~\ref{fig:linearity}) in each mode for each candidate value.

We generate a command matrix by taking the SVD of the interaction matrix, $M = USV^\intercal$, inverting and truncating the matrix of singular values $S$ (i.e. setting values past a certain cutoff to 0) to only include a user-specified number of modes to generate an inverse singular value matrix $S^\dagger$, and computing the pseudoinverse $C = VS^\dagger U^\intercal$. The number of control modes is varied on sky between 2 and 7 (see \S\ref{sec:cl_onsky_result}). We take as our reconstructed wavefront $C \cdot (\text{intensities} - \text{flat})$, where both `intensities' and `flat' are in the reduced and normalized WFS space. 

\begin{figure}
    \centering
    \includegraphics[width=\linewidth]{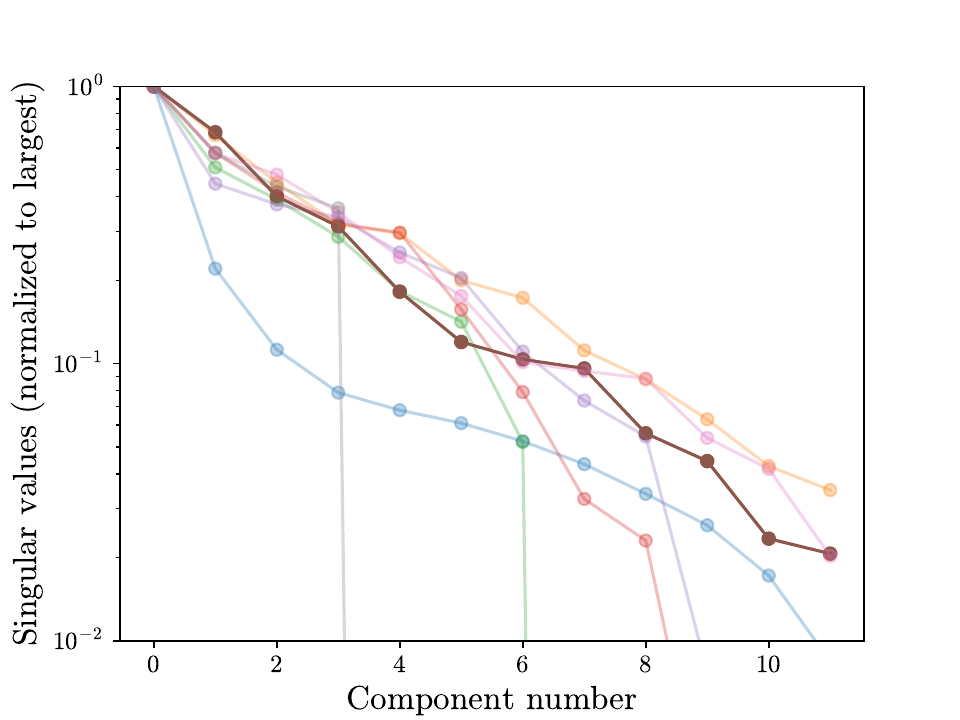}
    \caption{The singular value spectra from several interaction matrices taken on the afternoon and evening of 2025-08-07. The dark brown line is the interaction matrix used for on-sky control.}
    \label{fig:sv_spectra}
\end{figure}

Figure~\ref{fig:sv_spectra} shows the singular value spectra from several interaction matrices. The 10\%  singular value threshold used by \cite{Lin2025} is reached between 3 and 8 control modes depending on the alignment and on the poke amplitude used to build the interaction matrix. { Most interaction matrices that showed acceptable performance in off-sky testing had at least 6 controllable modes; some measured interaction matrices showed fewer modes, such as the blue and grey lines in Figure~\ref{fig:sv_spectra}. This is likely a result of suboptimal calibration parameters, such as the poke amplitude and the number of stacked frames at each DM position, which were empirically tuned. In all cases, we observed fewer modes than the expected number of around 15 (\S\ref{sec:pl}); this is likely a result of suboptimal coupling, as detailed in \S\ref{sec:optics}. In particular, the number of sensed modes is likely limited by the PSF being too large to allow features past $2\lambda/D$ to couple into the PL.}

\begin{figure}
    \centering
    \includegraphics[width=\linewidth]{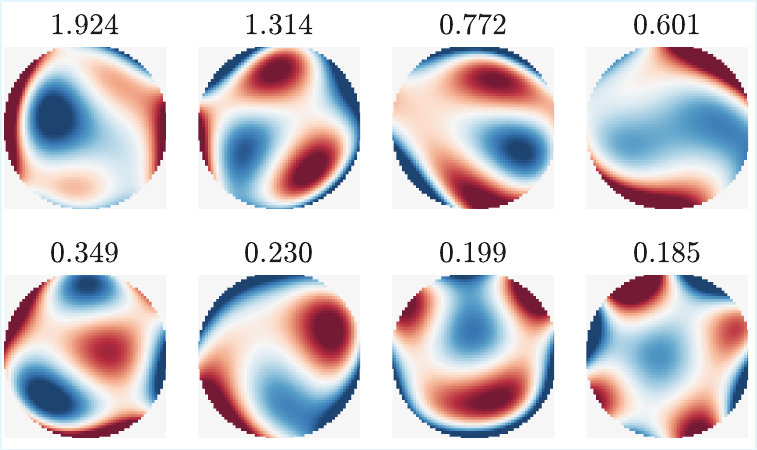}
    \caption{The control modes identified in the interaction matrix used for on-sky control, with the corresponding singular values.}
    \label{fig:control_modes}
\end{figure}

Figure~\ref{fig:control_modes} shows the control modes identified by taking the SVD of the interaction matrix used on sky. We note mostly low-order patterns but with significant edge features. We speculate that this is due to a static misalignment; this would cause PSF features past $2-3 \lambda/D$ that would otherwise fall outside the PL input to be seen, and these corresponding to high-order aberration modes for which most of the variation is on the edge of the pupil. However, confirming this is out of the scope of this work.

Figure~\ref{fig:linearity} shows linearity curves in the first three image sharpening modes. We observe linear ranges of approximately 13 nm in all modes but note that we underestimate the true aberrations; attempts at using a smaller calibration amplitude to rectify this did not show usable linear ranges due to noise sensitivity. We note that although RMS wavefront error differences are relatively small throughout this experiment, these correspond to large proportional shifts in image quality. This is due to poor initial image quality resulting from the alignment and coupling challenges described in \S\ref{sec:optics}.


We monitor performance using a custom-written live display on the control computer that shows the dark-subtracted PL camera readout, the current exposure time and gain on the PL camera, the number of frames being used per control iteration, the latest command sent from the image sharpening script to the DM, and the total WFE measured by the PL when a closed-loop control attempt is underway.

\section{Observations and results}

\subsection{Observing configuration}

\begin{figure}
    \centering
\includegraphics[width=\linewidth]{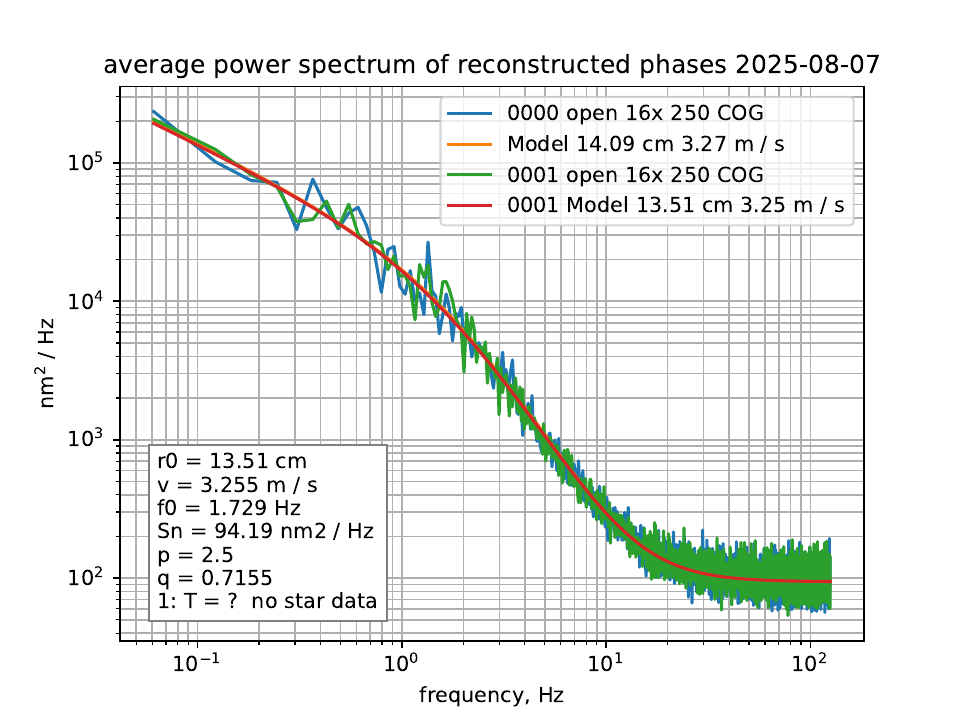}
    \caption{Open-loop power spectra of reconstructed phase screens from the ShaneAO Shack-Hartmann WFS, and the corresponding von K\'{a}rm\'{a}n profile fits, measured on 2025-08-08 UTC (2025-08-07/08 PDT).}
    \label{fig:telemetry}
\end{figure}

We observed Eltanin/$\gamma$ Dra ($m_H = -1.034$) with ShaneAO on 2025-08-08 UTC. We estimated seeing using the Shane telescope's guider camera and found it was 1.2''. We measured $r_0$ by fitting an open-loop power spectrum to a von K\'{a}rm\'{a}n profile, and found it was 13.51cm. Wind speeds and temperatures throughout the night were available due to the observatory's weather sensors; these recorded wind speeds of 0.4-3.6 m/s (consistent with the 3.2 m/s identified from the von K\'{a}rm\'{a}n fit), and outdoor temperatures of 21-23${}^\circ$C. Figure~\ref{fig:telemetry} shows the open-loop power spectrum used for the $r_0$ fit.

{ During all experiments, we initially close the first-stage ShaneAO loop and carry out an initial open-loop coupling optimization with respect to the PL camera using the spiral search method described in \S\ref{sec:coupling}. Given the large static aberrations on our PSF, performance metrics for first-stage correction (such as Strehl) would require images from ShARCS. However, our setup picks off light that would otherwise go to ShARCS, so these data are unavailable.}


\subsection{Closed-loop control with the PL}
\label{sec:cl_onsky_result}

\begin{figure}
    \centering
    \includegraphics[width=\linewidth]{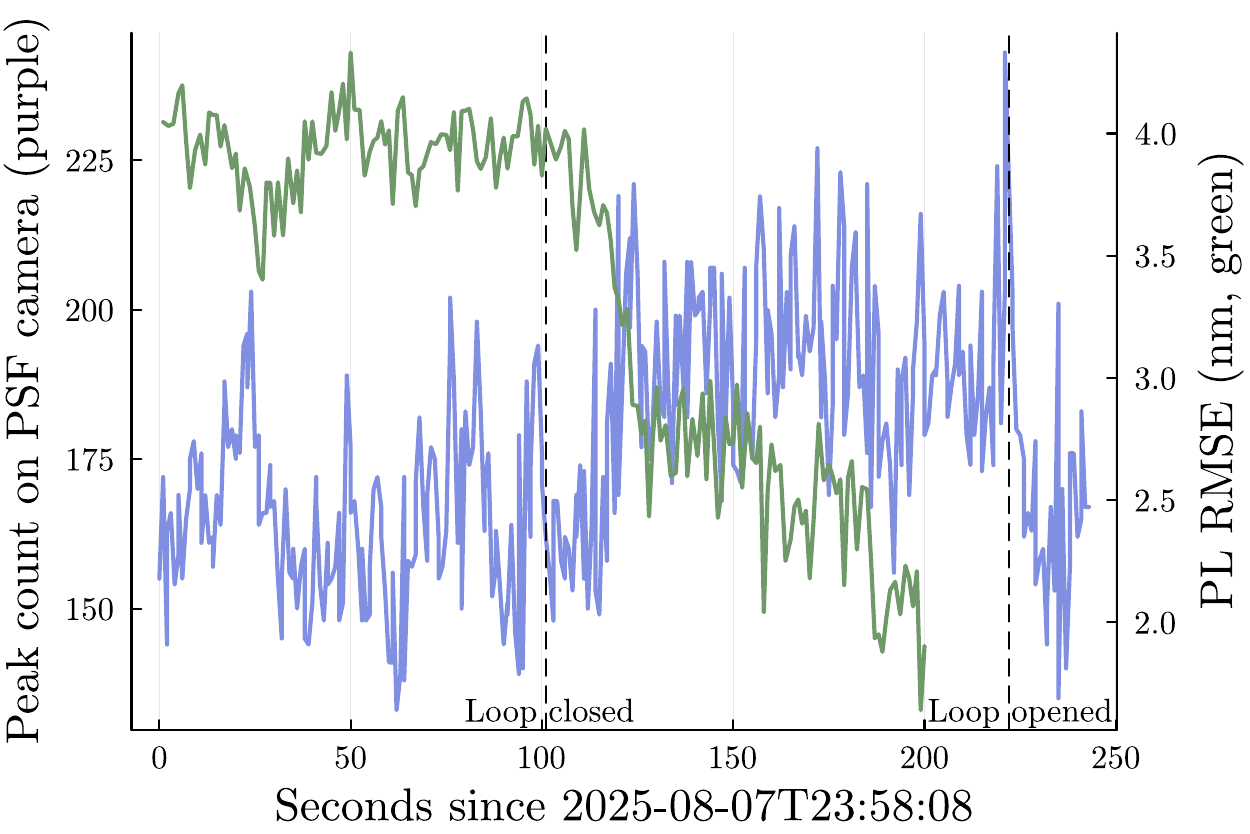}
    \caption{Closed-loop control improves the PSF (purple) and reduces measured wavefront error (green) on the PL at the same time.}
    \label{fig:closed_loop}
\end{figure}

Figure~\ref{fig:closed_loop} shows an example PL closed-loop control run. We observe improved PSF quality together with a reduction in measured wavefront error on the PL. Since the main aberrations we are correcting are quasi-static, the system state does not substantially change immediately when we stop applying DM commands; for comparison the system is therefore reset to the best flat as identified in \S\ref{sec:coupling} at the end of each PL closed-loop attempt. The PL controller is a leaky integrator with gain = 0.3 and leak = 0.95. We stack 3 frames on the WFS camera (at 1s exposures) per DM update. 

The RMS wavefront error as measured by the PL reduces from around 3.8nm in open-loop to 1.3nm in closed-loop. Note that these measurements do not reflect the total wavefront error seen either at the PL or the PSF camera; they pertain to the part that can be corrected using the image sharpening script. { In particular, since the bulk of the error is static and is present during calibrations as well as on sky, the dynamic range of the PL is not a limiting factor. Since the PL reconstruction is linearized around the best flat position from off-sky calibrations, rather than around a diffraction-limited beam, it is only relevant to consider aberrations that are different off-sky vs. on-sky. Relative to the off-sky alignment, the observed quasi-static non-common-path aberrations have comparable magnitudes to what the lantern can sense, and to what the image sharpening script is able to correct, independent of the high degree of static error.} Despite the relatively small magnitude of the measured wavefront errors, this amount of change is sufficient to show large relative changes in image quality, { even if the absolute change may be small as it is limited by static aberrations.}

\begin{figure}
    \centering
    \begin{subfigure}{0.5\linewidth}
        \centering
        \includegraphics[width=\linewidth]
        {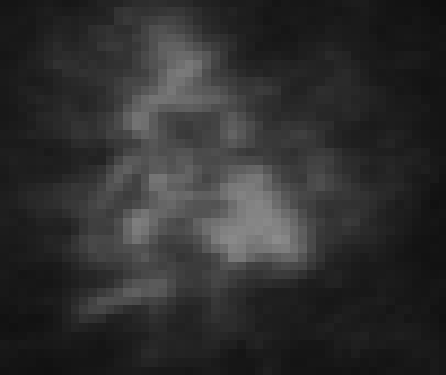}
        \caption{Open-loop PSF, 23:59:11} \label{fig:ol_psf}
    \end{subfigure}\hspace*{\fill}
    \begin{subfigure}{0.5\linewidth}
        \centering
        \includegraphics[width=\linewidth]{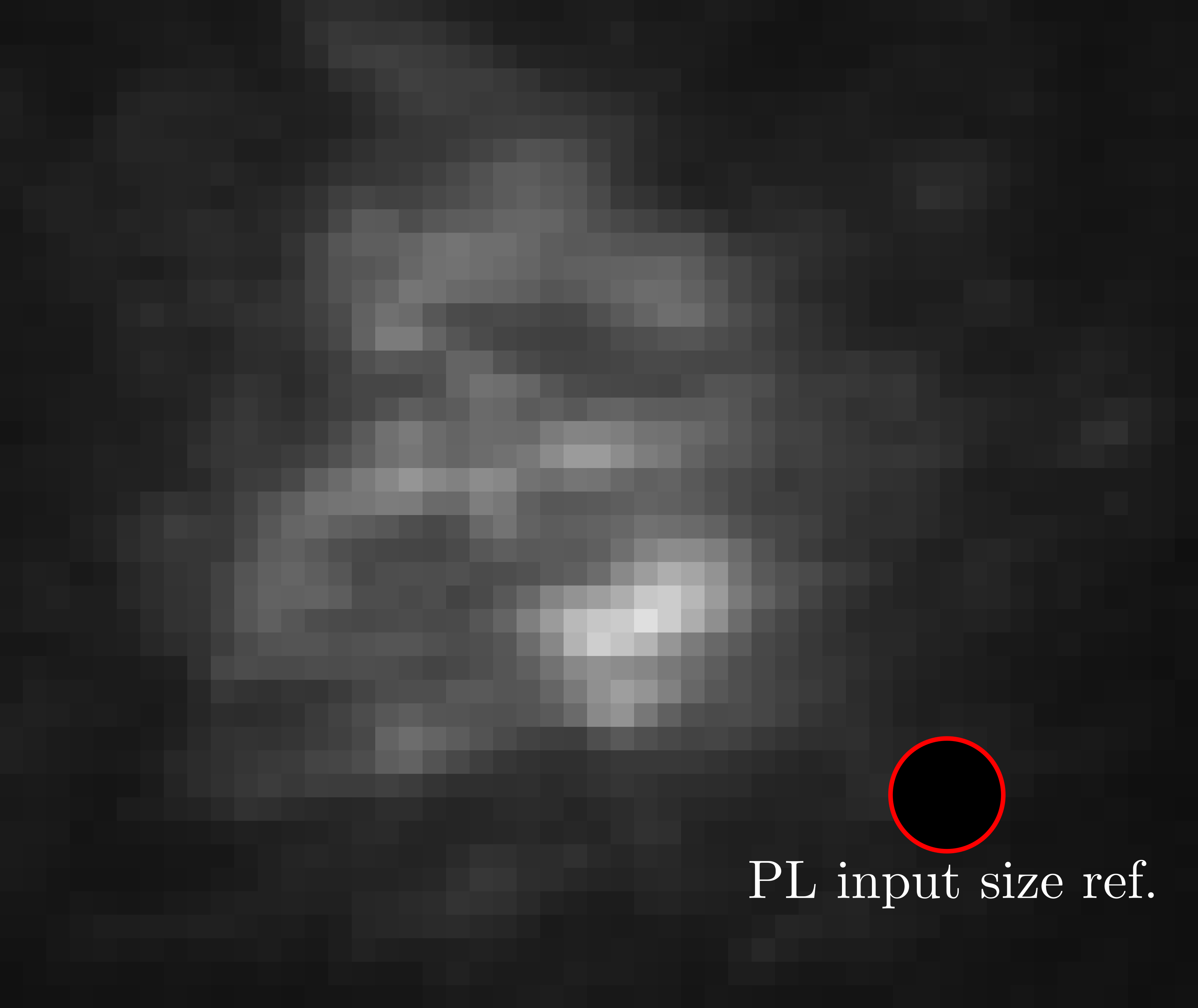}
        \caption{Closed-loop PSF, 00:01:50}
        \label{fig:cl_psf}
    \end{subfigure}
    \caption{Open- and closed-loop PSFs corresponding to the closed-loop run in Figure~\ref{fig:closed_loop}. The closed-loop PSF image includes a size reference for the PL input. Note that the location of this size reference is arbitrary.}
    \label{fig:cameras}
\end{figure}

\begin{figure}
    \centering
    \includegraphics[width=\linewidth]{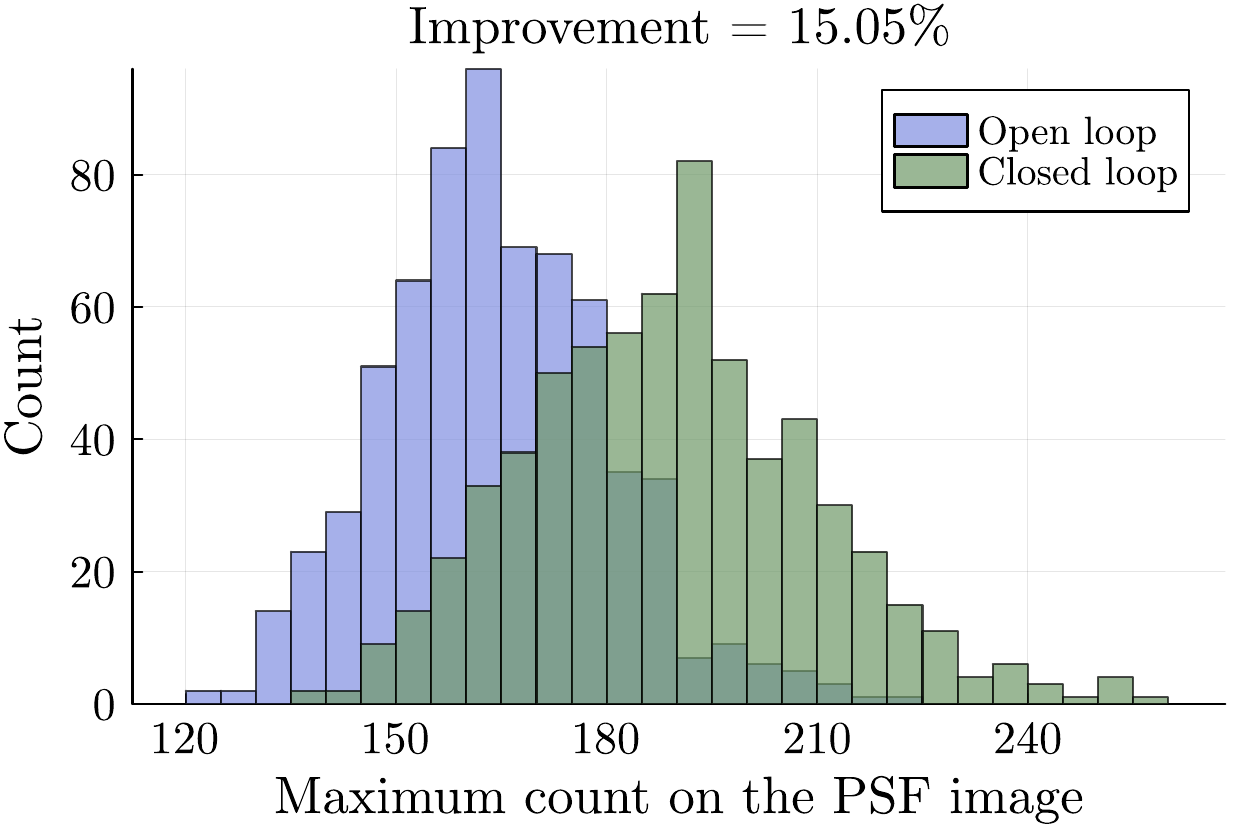}
    \caption{The average peak count on the PL camera is higher in closed loop than in open loop.}
    \label{fig:imagecount}
\end{figure}

We successfully closed the control loop 9 times, as determined by the RMSE measured on the PL being significantly lower than when no commands were applied. We confirm that this control led to improved image quality by assessing the peak count on the PSF camera when in open loop compared to closed loop. A representative pair of PSFs is shown in Figure~\ref{fig:cameras}, and an overall comparison of the peak count across all 9 runs is shown in Figure~\ref{fig:imagecount}. Figure~\ref{fig:cl_psf} also includes a size reference for the entrance/multi-mode end of the photonic lantern. We expect to only be sensitive to PSF features around the peak that would fall within this; we do not sense the diffuse outer parts of the PSF. We observed intensity increases primarily over this size scale, as expected.

\section{Conclusion and Future Work}
\label{sec:conclusion_future}
We have demonstrated on sky that undispersed, narrow-band photonic lantern outputs can be used for wavefront sensing in a second-stage adaptive optics control configuration, providing an improvement in final residual wavefront error and image quality. Further, we have demonstrated that narrow-band spectral information from a 19-port PL is sufficient to correct a substantial component of quasi-static non-common-path aberrations. This enables wavefront control at a single wavelength, concurrently with science imaging that makes use of the remaining parts of the spectrum.

Substantial performance improvements will be possible with a more stable and better-aligned optical configuration, as is intended with the APALO test bench. The current performance ceiling is set by the high degree of static aberration at the best flat position, as can be seen in Figure~\ref{fig:psfcam}. As mentioned in \S\ref{sec:optics}, this is largely due to limitations of the prototype PLIU assembly. All of the lenses in the PLIU are singlets and coated for 600-1000 nm, as previous experiments in this configuration were done at visible wavelengths \citep{DeMartino2024}. This presents a further source of static aberrations in our experiment; a dedicated setup for wavefront sensing at 1550 nm could achieve better PSF quality as well as better coupling into the PL.


We have observed a factor of $\sim$5 difference in the PL throughput depending on the quality of alignment day to day, which affects the exposure time and therefore the loop rate. Further, optimal coupling would create a more even distribution of light across the ports. In this experiment, the effective number of ports that can be used for reliable wavefront reconstruction was limited by this unevenness; in several cases at the flat wavefront position, we observed detector saturation on some ports while no light was visible through others, and image sharpening was not always sufficient to prevent this. Even during successful closed-loop operation on sky, the detector saturated on at least one port in all frames owing to the high dark counts (as detailed in \S\ref{sec:software}). Further, no light was recorded through at least one port at some positions during calibrations and on sky. In addition to improved alignment at the flat position, this could be further remedied by second-stage tip/tilt control, considering the drastic difference in PL throughput when the PSF is offset in x/y.

In this work, we employed a dual-wavefront sensor single-conjugate AO configuration, in which two WFS signals are combined to control the same DM. Since the PL was responsible only for static/quasi-static aberrations, WFS reference offsetting was sufficient to prevent any conflicts between the two measurements. However, if we were to extend to faster-varying atmospheric residuals, this strategy would result in two different control objectives (as a result of the two WFSs steering to different points set by the different NCPAs on both imaging arms) and in control loop instability or suboptimal performance. We would therefore require a customized control law such as that demonstrated by \cite{Gerard2023}, which filtered the reconstructions from both WFSs to avoid temporal overlap.

Future work could involve implementing nonlinear wavefront reconstruction in order to reduce modal crosstalk and extend dynamic ranges; a nonlinear reconstruction that could be empirically calibrated may also be more robust to misalignments. \cite{Lin_Fitzgerald_2024} explored several nonlinear reconstruction methods that would be suitable for the PL, such as fitting the intensity-to-phase map with radial basis functions. However, the number of sample points required to calibrate these methods for a 19-port PL is too high considering our current DM latency. Direct access to the real-time controller, together with improved coupling optics to allow calibrations at shorter exposure times, may remedy this. Transfer matrix identification for an $N$-port PL is possible within $N^2$ measurements \citep{Sengupta2024,Romer2025} (in our case 361), which is feasible with the current latency and may enable other nonlinear reconstructors (e.g. \citealt{Haffert2024}). However, this would require better optical stability than the current configuration, in order to ensure that results from this identification would represent what we would observe on sky.

\begin{acknowledgments}
    We thank the anonymous referee for their helpful comments, which have greatly improved the quality and clarity of this work. We thank Paul Lynam and Jon Rees for supporting observations; William Deich and JS Roark for custom upgrades at ShaneAO to support this experiment; Shawn Stone, Paul Canton, Dan Espinosa, Matt Brooks, Norm Jackson, and Donnie Redel for operating the Shane telescope; and the custodial staff at Lick Observatory for facilitating our stays at the summit. We thank Jonathan Lin, Michael Fitzgerald, Emiel Por, Vincent Chambouleyron, Dominic Sanchez, Eden McEwen, and Parth Nobel for valuable conversations. This work was funded in part by the Heising-Simons Foundation Grant 2020-1822. JD acknowledges support from the Cota-Robles Fellowship. UCSC co-authors acknowledge support from the UCSC Office of Research, the UCSC Division of Physical and Biological Sciences, and the University of California Observatories. UCF co-authors acknowledge support from Air Force Office of Scientific Research under award number FA9550-24-1-0332, the UCF SPICE AEP initiative, and the UCF PHAST Jumpstart initiative.
\end{acknowledgments}

\begin{contribution}
    AS led the experimental design, wrote the image reduction and control pipeline, conducted the experiments, and led the data analysis and interpretation. JD led the optical installation and alignment. MDM led the optical design and consulted on laboratory testing. RJC contributed to experimental design and data analysis.  SC and MS contributed to creating and installing the image processing software. EG, KB, DD, and PH contributed to optical design and integration with the Shane telescope. NS and OG contributed to creating and validating the control pipeline. TC, CD, SE, RA-C and SY manufactured the photonic lantern. All authors contributed to writing the manuscript.
\end{contribution}

\facilities{Shane (ShARCS)}

\software{dao \citep{dao}, paramiko \citep{paramiko}, numpy \citep{numpy}, matplotlib \citep{matplotlib}, Plots.jl \citep{plotsjl}}

\appendix

\section{Calibration for ShaneAO image sharpening}
\label{sec:image_cal}
In this work, we interacted with the ShaneAO deformable mirrors using a script called \textit{imageSharpen}. This is primarily intended for observers using ShARCS to refine images, by moving sliders for each low/mid-order Zernike mode and assessing the image quality either by eye or via ShaneAO's Strehl ratio calculator. The magnitude of the commands sent using \textit{imageSharpen} uses an arbitrary scale from -1 to 1 (henceforth referred to as $\textrm{i.u.}$ for `\textit{imageSharpen} units'), intended to map to a reasonable range for observers to discern differences by eye without applying sufficiently large offsets to interfere with loop stability. In order to interpret the results of this experiment in terms of standard units, it was necessary to empirically identify the conversion factor from i.u. to nanometers.

With ShARCS in its usual position, after having gone through its usual alignment procedure, and without any of the custom optics used in this work but with ShaneAO's Brackett $\gamma$ filter ($\lambda = 2.165 \mu$m) in the path, we sent offsets to the Shack-Hartmann using the same method as in the on-sky experiment. We used ShaneAO's Strehl ratio calculator to measure the Strehl ratio as a function of the offsets in each Zernike mode, in a range from -0.4 to +0.4 i.u. with a step of 0.05. We fit these curves to the Mar\'echal approximation relating Strehl ratio (relative to the Strehl for a flat wavefront, which was 0.7) to total wavefront error, in the following form:

\begin{align*}
    SR(a) = 1 - 4\pi^2 \left(c (a - a_0)\right)^2
\end{align*}

where $a$ is the input in i.u., $a_0$ is an offset term also in i.u., and $c$ is a conversion factor from i.u. to radians. We have $a$ and $SR(a)$ as our data and we fit $c$ and $a_0$ for each Zernike mode.

\begin{figure}
    \centering
    \includegraphics[width=0.8\linewidth]{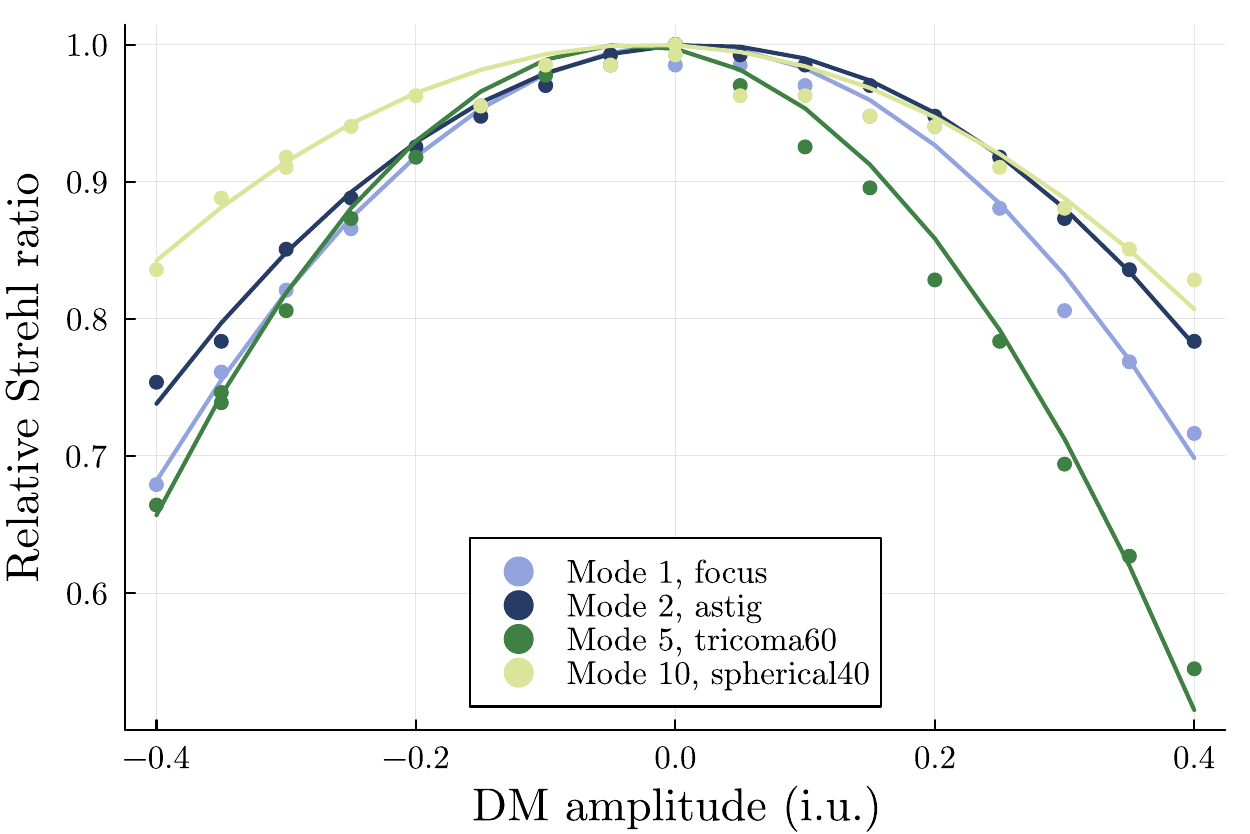}
    \caption{The conversion from \textit{imageSharpen} units to nanometers was constrained by fitting data from ShARCS to the Mar\'{e}chal approximation.}
    \label{fig:dm_calibration}
\end{figure}

Figure~\ref{fig:dm_calibration} shows four representative curves of the twelve considered, each at a different Zernike order. We find that $c$ ranges from 0.127 to 0.267 radians per i.u., with the lower values corresponding to the five 4th-order modes. We neglect mode-to-mode differences and consider only the mean value of 0.184 radians per i.u., which corresponds to $0.184 / (2\pi) \times 2165 \text{nm} = 63.4$ nm per i.u. All data in this experiment that involve image sharpening were collected in i.u. and converted to nm using this calibration.

{ To validate this unit conversion and its correspondence to our observed improvement in image quality, we ran a simulation in which we chose an initial phase screen by drawing a random linear combination of focus/astigmatism/astigmatism45. We then subtracted 14 nm of wavefront error from this phase screen, corresponding to the variation on the DM via image sharpening in the closed-loop run in Figure~\ref{fig:closed_loop} assuming that our unit conversion from i.u. to nm is correct. We simulated PSFs corresponding to both phase screens and divided the resulting Strehl ratios to get an expected improvement fraction. Over 1000 realizations of this process, we found that for an average initial Strehl ratio of $0.029 \pm 0.009$, the improvement is $14 \pm 6$\%. Our improvement of 15\% in this work therefore closely matches the expected improvement considering the initial image quality.}

\bibliographystyle{aasjournal}
\bibliography{onsky}

\end{document}